\documentclass[10pt,a4paper]{article}
\pdfoutput=1
\usepackage[english]{babel}  
\usepackage{longtable,subfigure}
\usepackage{amsmath,amsfonts,amssymb,bm}\interdisplaylinepenalty=2500
\usepackage[T1]{fontenc}
\usepackage{graphicx,color,rotating,epstopdf}
\graphicspath{{figs-pdf/}}
\def\PrintGraphicFileName{1}			
\newcommand{\namedgraphics}[3]{
	\parbox{#3}{%
	\ifnum\PrintGraphicFileName>0\rotatebox{90}{\smash{\ttfamily\scriptsize\raisebox{0.8em}{#2}}}\fi%
	\hspace*{\fill}\includegraphics[scale=#1]{#2}\hspace*{\fill}}}

\newcommand{\ToDay}{~~~---~~~Jan.\ 31, 2012}
\newcommand{\TODAY}{January 31, 2012}
\title{\boldmath Phase Noise in RF and Microwave Amplifiers}
\author{Rodolphe Boudot and Enrico Rubiola\\
\small web page \texttt{http://rubiola.org}
\\[4em]\includegraphics[width=0.35\textwidth]{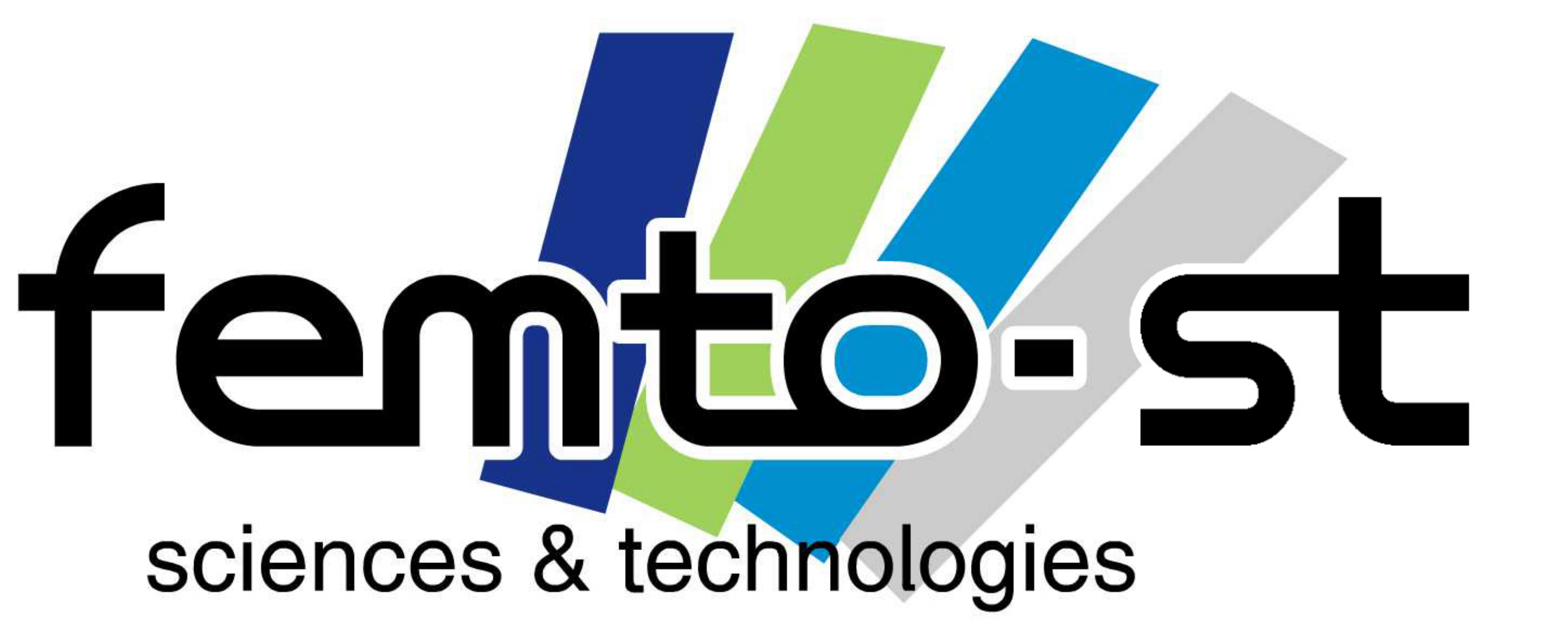}\\[0.5em]
\small FEMTO-ST Institute\\[-0.5ex]
\small CNRS and Universit\'e de Franche Comt\'e, 
\small Besan\c{c}on, France\\[1.5em]}
\date{\small\TODAY}
\def\FigOneScale{0.8}
\def\FigTwoScale{0.64}
\def\SchemeScale{0.64}
\def\SpectraScale{0.64}
\def\OeoSpectrumScale{0.8}
\def\GeneralizedModelScale{0.60}
\def\SimulationScale{0.88}

\begin{document}
\maketitle

\begin{abstract}
Understanding the amplifier phase noise is a critical issue in numerous fields of engineering and physics, like oscillators, frequency synthesis, telecommunications, radars, and spectroscopy; in the emerging domain of microwave photonics; and in more exotic fields like radio astronomy, particle accelerators, etc.

This article analyzes the two main types of phase noise in amplifiers, white and flicker. So, the power spectral density of the random phase $\varphi(t)$ is $S_\varphi(f)=b_0+b_{-1}/f$.  White phase noise results from adding white noise to the RF spectrum in the carrier region.  For a given RF noise level, $b_0$ is proportional to the inverse of the carrier power $P_0$.  By contrast, $b_{-1}$ is a parameter of the amplifier, constant in a wide range of $P_0$.  The consequences are the following.  Connecting $m$ equal amplifiers in parallel, $b_{-1}$ is $1/m$ times that of one device.  Cascading $m$ equal amplifiers, $b_{-1}$ is $m$ times that of one amplifier.    Recirculating the signal in an amplifier so that the gain increases by a power of $m$ (a factor of $m$ in dB) due to positive feedback (regeneration), we find that $b_{-1}$ is $m^2$ times that of the amplifier alone.  The feedforward amplifier exhibits extremely low $b_{-1}$ thanks to the fact that the carrier is ideally nulled at the input of its internal error amplifier. 

Starting from the fact that near-dc flicker exists in all electronic devices, even if generally not accessible from outside, the simplest model for phase flickering is that the near-dc $1/f$ noise modulates the carrier through some parametric effect in the semiconductor.  This model predicts the behavior of the (simple) amplifier and of the different amplifier topologies.   Numerous measurements on amplifiers from different technologies, also including some old samples, and in a wide frequency range (HF to microwaves), validate the theory.  In turn, our results provide design guidelines and suggestions for improved CAD simulations.
\end{abstract}

\clearpage
\tableofcontents
\clearpage

\section{Introduction}\label{sec:amx-introduction}
Low phase noise amplification is crucial in a variety of applications. In the oscillator, the phase noise of the sustaining amplifier is converted into frequency noise via the Leeson effect \cite{Leeson1966pieee, Rubiola-2008-Cambridge--Leeson-effect, Nallatamby2003mtt-Leeson, Sauvage1977im-Leeson}.  Hence the oscillator phase fluctuation, which is the integral of frequency, diverges in
the long run.  In turn, the oscillator noise impacts on the bit error rate \cite{Nezami1998mrfm,Tomba1998tc} and on security \cite{Howe2005fcs} in communications, and on radars \cite{Krieger2006grsl-noise-radar,Scheer1990radar-conf}.  Doppler and chirp radars require ultra-low phase noise to avoid that the oscillator noise sidebands exceed the echo signal.  Low phase noise
amplification is important in precise synchronization systems because phase represents time.  Finally, the books \cite{Robins:phase-noise,Kroupa:frequency-stability} provide useful overview, though not up to date.

Near-dc $1/f$ noise, discovered in the 1930s \cite{Christiansen1936bstj-flicker}, is now considered an ubiquitous phenomenon for which no generally-agreed unification available.  Most most models for electronic components resort to two original articles \cite{Hooge69pla, McWhorter1957ssf}.  
Phase flickering can only originate from near-dc $1/f$ noise brought to the vicinity of the carrier.  This occurs because in the absence of a carrier, the noise at the amplifier output is nearly white.  Since the near-dc flicker is generally stationary, $1/f$ phase noise is cyclostationary.

The problem with non-linear noise modeling is that the model rely on the identification of the near-dc noise sources, which can in turn be non-linear or associated to a non-linear
circuit element \cite{Boudot-2006-ELL--Low-noise-oscillator, Siweris1986emc, Llopis2001mtts}.  Since the conversion of near-dc noise into phase noise is generally not implemented in CAD programs, the simulation may require dedicated software.  Although these models are not a perfect representation of the device physics, some of them provide results in quite a reasonable agreement with the measured phase noise \cite{Llopis2001mtts, Cibiel-2004-UFFC--sapphire, Gribaldo2005eftf}.
Some theoretical models, supported by experiments, provide useful information about amplifier $1/f$ phase noise for several technologies \cite{Pucel1983mtts, Rhodin1984mtts, Walls-1997-UFFC, Ferre-Pikal-1997-UFFC, Ferre-Pikal-2008-UFFC}.  Conversely, the more accurate semiconductor-physics approach \cite{Bonani-2002-ED} and the related microscopic models are complex and difficult to use.

To conclude, the amplifier phase noise is more understood, albeit the pieces of information are scattered in many articles.  By contrast, little information is available about the consequences of these mechanisms, and on more complex amplifier architectures.  This article is intended to fill this gap, providing insight, practical knowledge, design rules and extensive experimental confirmation.

\section{Phase Noise Mechanisms}\label{sec:PNMech}
\begin{figure}
\centering\includegraphics[scale=\FigOneScale]{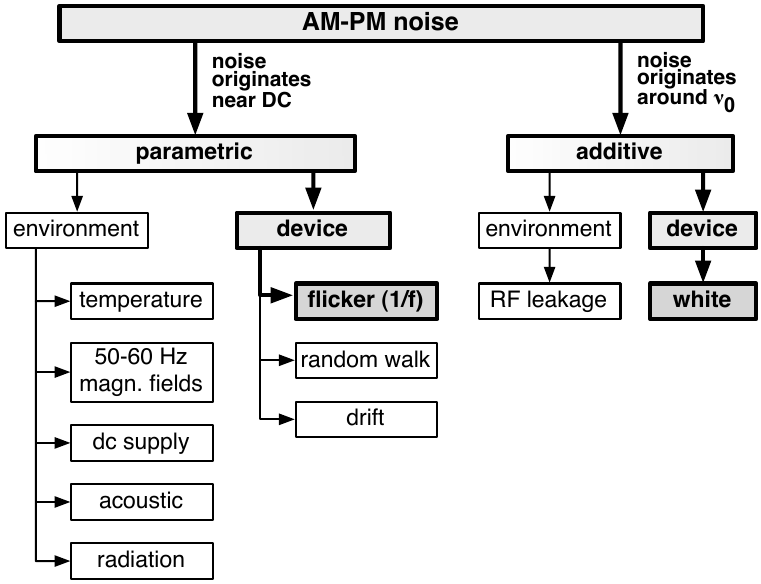}
\caption{Amplifier internal phase noise mechanisms.}
\label{fig:NoiseMechanisms}
\end{figure}

Figure~\ref{fig:NoiseMechanisms} presents a rather general  panorama of noise in amplifiers, suggested by experience and physical insight.  In this article we restrict the attention to white and flicker noise.  The reason is that among the noise types originated from the amplifier \emph{inside}, white and flicker are those responsible for short-term phase noise.  So, the phase noise spectrum is completely described by the first two terms of the polynomial law
\begin{equation}
S_\varphi (f) = b_0 + \frac{b_{-1}}{f}~. \label{eq:specAmpli}
\end{equation}
The white phase noise $b_0$ derives by adding to the carrier a
random noise of power spectral density $N=FkT_0$, where $k$ is the
Boltzmann constant, and $F$ is the amplifier noise figure defined at
the reference temperature $T_0=290$ K (17 $^\circ$C). It is useful
to have on hand the following numerical values
\begin{equation*}
kT_0 = 4{\times}10^{-21}~\mathrm{J~~(-174~dBm/Hz)}.
\end{equation*}
In modern low-noise amplifier, $F$ is typically of 0.5--2 dB.  It may
depend on bandwidth, on the loss of the input impedance-matching
network, and on technology. If the actual  temperature is not close
enough to $T_0$, the quantity $F$ is meaningless.  In this case, the noise is
described by $N=kT_e$, where $T_e$ is the equivalent noise
temperature, which includes amplifier and its input termination. We
assume that $N$ is independent of frequency in a wide range around
the carrier frequency $\nu_0$, as it happens in most practical
cases.

Adding $N$ to a carrier of power $P_0$ results in random phase
modulation of power spectral density
\begin{equation}
b_{0} = \frac{FkT_{0}}{P_0}~.
\label{eq:b0}
\end{equation}
The above holds in the linear region of the amplifier.  If the
amplifier is operated in large-signal regime, where it is nonlinear
or saturated, $F$ may increase \cite{Cibiel2004mtt-Si-transistors,
Ivanov2000uffc}.

At low frequencies, the amplifier phase noise is of the $1/f$ type, which currently referred to as flicker.  Near-dc flicker noise takes place at the microscopic scale \cite{Hooge69pla,McWhorter1957ssf},
for little or no correlation is expected between different region of the device.   This is supported by the fact that the probability density function is normal \cite{Brophy-1968-PR--flicker-statistics}.  Such distribution originates from the central-limit theorem in the presence of a large population of independent phenomena. 

\begin{figure}
\centering\includegraphics[scale=\FigTwoScale]{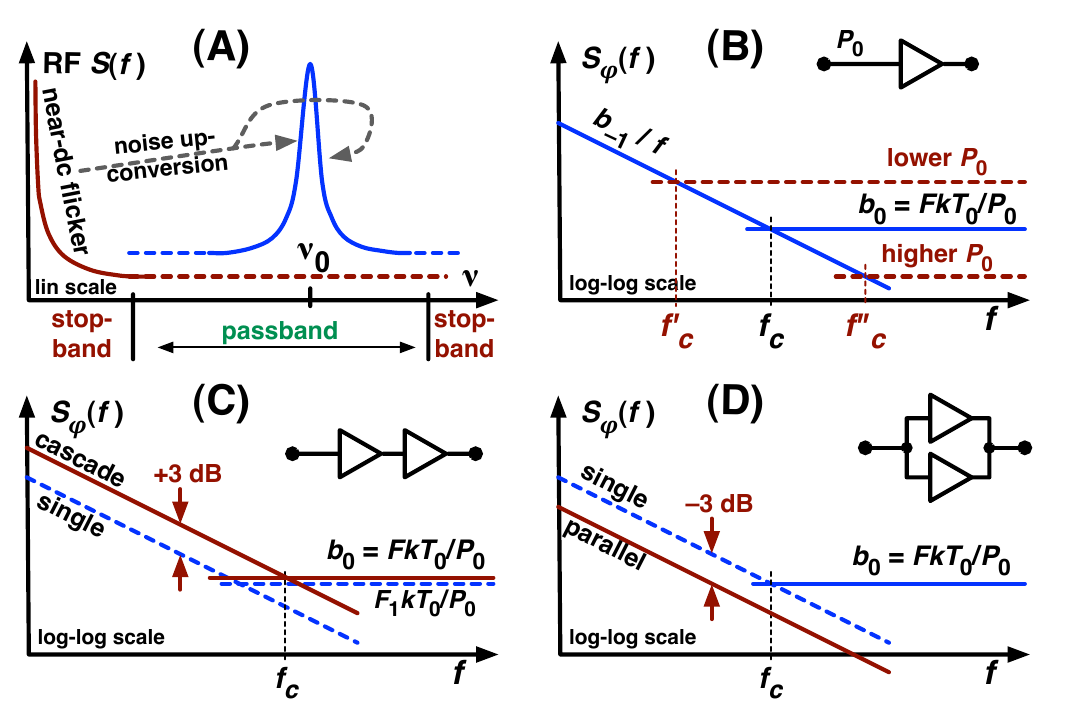}
\caption{Phase noise rules for several amplifier topologies. 
(A): noise up-conversion from near-dc the carrier frequency, which originates $1/f$ phase noise. 
(B): single amplifier. (C): cascaded amplifiers. (D): parallel amplifiers.}
\label{fig:4schemes}
\end{figure}

Understanding phase flickering in amplifiers starts from the simple fact that noise is white in the absence of a carrier.  Besides the experimental evidence, the heuristic proof given by Nyquist \cite{Nyquist-1928-PR} for thermal noise is convincing also after introducing the noise figure $F$, which is not necessarily a thermal phenomenon.
Close-in noise shows up only when the carrier is sent at the input.
This means that phase flickering can only originate from
up-conversion of the near-dc $1/f$ noise, as shown in
Fig.~\ref{fig:4schemes}~A. The noise up-conversion can be described
as follows.  We denote with $u(t)=U_0e^{j2\pi\nu_0t}+n'(t)+jn''(t)$
the input signal, where $U_0e^{j2\pi\nu_0t}$ is the `true'
(accessible) input and $n=n'+jn''$ the near-dc equivalent noise at
the amplifier input; and with $v(t)=a_1u(t)+a_2u^2(t) +
\textit{noise}$ the output signal. The near-dc noise $n(t)$ is
\emph{not} the random signal that would ideally be measured with an
oscilloscope.  Instead, it is an abstract quantity with spectrum
proportional to $1/f$ that accounts for the parametric nature of
flicker. The amplifier is described as a (smooth) nonlinear function
truncated at the second order, where the coefficient $a_1$ is the
(usual) voltage gain denoted with $A$ elsewhere in this article.
Expanding $v(t)$ and selecting only the $2\pi\nu_0$ terms we get
\begin{align}
v(t)=a_1U_0e^{j2\pi\nu_0t} + 2a_2[n'+jn'']U_0e^{j2\pi\nu_0t}~,
\label{eq:v-t}
\end{align}
from which
\begin{align}
\alpha(t)&=2\frac{a_2}{a_1}n'(t) &
S_\alpha(f)&=4\frac{a^2_2}{a^2_1}S_{n'}(f)
\label{eq:alpha-t}\\[1ex]
\varphi(t)&=2\frac{a_2}{a_1}n''(t)&S_\varphi(f)&=4\frac{a^2_2}{a^2_1}S_{n''}(f)
\label{eq:phi-t}~.
\end{align}
Equations (\ref{eq:v-t}), (\ref{eq:alpha-t}) and (\ref{eq:phi-t}) express
the simple fact that the noise sidebands are proportional to the
carrier amplitude, and therefore AM and PM noise are independent of
the carrier amplitude or power.   In this representation we use the
nonlinearity, present in virtually all devices, to transpose the
random signal $n(t)$.  Of course a fully-parametric model yields the
same results, at a cost of heavier formalism.

Experiments show that $b_{-1}$ is almost independent of the carrier power \cite{Gribaldo2005eftf,
Walls-1997-UFFC, Halford1968fcs, Hati2003fcs} if the amplifier operates in linear regime or in mild compression.  The quasi-static
perturbation technique provides fairly good agreement between
simulated and experimental $1/f$ phase noise data in Silicon and SiGe
amplifiers \cite{Cibiel-2004-UFFC--sapphire}. Other investigations
describe the $1/f$ phase noise as a modulation from the near-dc
$1/f$ current fluctuation in microwave HBT amplifiers
\cite{Ferre-Pikal-2008-UFFC} and in InGaP/GaAs HBTs
\cite{Chambon2007mtt}. The analysis of the literature cited
indicates that, regardless of the theoretical approach and of the
amplifier technology, the amplifier \emph{behavior} is that of a
linear phase modulator driven by a near-dc process
\begin{equation}
b_{-1} = C \qquad\text{(constant, independent of $P_0$)}~.
\label{eq:b-1}
\end{equation}
Neither the near-dc noise nor the modulation efficiency are affected
by the carrier power, unless the amplifier is pushed in strong compression.  If this happens, the dc bias changes.  In
turn, small changes of $b_{-1}$ are expected in an unpredictable way.
Our experiments, detailed in Sec.~\ref{sec:ExpProof} confirm this
behavioral model.

\section{Analysis and design rules}\label{sec:Rules}
\subsection{Single Amplifier}
The typical phase-noise pattern found in amplifier is shown in Fig.~\ref{fig:4schemes}\,B\@.  
An amazing fact comes immediately from (\ref{eq:b0}) and (\ref{eq:b-1}), that the corner frequency is given by
\begin{equation}
f_c = \frac{b_{-1}}{FkT_0}\:P_0~.
\label{eq:f-corner}
\end{equation}
This fact has been succesfully used to reverse-engineer the oscillators from their noise, identifying
some relevant parameters like the resonator $Q$ and driving power \cite[Chap.\,6]{Rubiola-2008-Cambridge--Leeson-effect}, \cite{Rubiola-2007-UFFC--Xtal-flicker}.

It is worth pointing out that the flicker corner frequency $f_c$ sometimes found in the amplifier specifications is misleading because it is presented as a parameter of the amplifier, as it was rather constant, at least in `normal' operating range.  
In SPICE and in some other CAD programs the flicker is described by $f_c$, introduced as a fixed parameter in the device model.  This is an unfortunate choice for the same reason.  
Replacing the parameter ``$f_c$'' with (\ref{eq:f-corner}) would result in improved usability.

\subsection{Cascaded Amplifiers}
When several amplifiers are cascaded (Fig.~\ref{fig:4schemes}\,C), the noise figure of the chain is given by the Friis formula \cite{Friis1944ire}
\begin{equation}
F = F_1 + \frac{F_2 - 1}{A_1^2} + \frac{F_3 -1}{A_1 ^2 A_2^2} +
\frac{F_4 -1}{A_1^2 A_2^2A_3^2} +\ldots~, \label{eq:Feq}
\end{equation}
where $A$ is the voltage gain. The Friis formula expresses the fact that the noise of the first stage is $F_1kT_0$, including the input termination, and the noise $(F_i-1)kT_0$ of the $i$-th stage ($i\ge2$) is referred to the input after dividing by the power gain of the $i-1$ preceding stages.  By virtue of (\ref{eq:b0}), the obvious extension of the Friis formula to phase noise is
\begin{equation}
b_0 = \left[F_1 + \frac{F_2 - 1}{A_1^2} + \frac{F_3 -1}{A_1 ^2
A_2^2} +  \frac{F_4 -1}{A_1^2 A_2^2A_3^2} +\ldots\right] \frac{k
T_0}{P_0}~. \label{eq:b0chain}
\end{equation}
In most practical cases the noise of the chain is chiefly determined
by the noise of the first stage.  This applies to the RF
spectrum, and also to the phase noise spectrum.

By contrast, the flicker phase noise is ruled by (\ref{eq:b-1}).
Since the amplifier $1/f$ phase noise processes in different devices are statistically independent
and also independent of the carrier power, the $1/f$ noise of a chain of $m$ amplifiers is
\begin{equation}
b_{-1} = \sum_{i=1}^m (b_{-1})_i~.
\label{eq:b-1ChainCasc}
\end{equation}
Cascading two (three) equal amplifiers, the phase flicker is 3 dB (4.8 dB) higher than that of the single one.

Combining white noise (\ref{eq:b0chain}) and flicker noise (\ref{eq:b-1ChainCasc}), we find the spectrum shown in Fig.~\ref{fig:4schemes}\,C.

\subsection{Parallel Amplifiers}
A \emph{parallel amplifier} (PA) as an amplifier network in which $m$ amplifier cells of the same gain share equally the burden of delivering the desired output power.  Several configurations are possible.  The push-pull configuration uses $180^\circ$ junctions, which suppresses the even-order harmonic
distortion, appreciated in audio applications.  The balanced amplifier \cite{Pozar:microwave-engineering:3ed} uses $90^\circ$ junction to improve input and output impedance matching.  The distributed amplifier \cite{Pozar:microwave-engineering:3ed}, preferred when a wide frequency range is to be achieved at any cost, uses a series of taps in a delay line to put the cells at work.

For the sake of analysis simplification, we assume that
\begin{itemize}
\item the cells are equal, and have voltage gain $A$, input and output impedance $R_0$ and noise figure $F$,
\item the input power-splitter and the output power-combiner are loss-free\footnote{In the case of the distributed amplifiers, it is conceptually impossible that all cells handle the same power.  Yet this hypothesis helps to understand.} and impedance matched to $R_0$.
\end{itemize}
Accordingly, the gain is equal to the gain $A$ of a cell, and the compression power is $m$ times the compression power of one cell.  

Denoting with $P_0$ the input power, the power at the input of each cell is $P_0/m$.  Consequently the white phase noise is
\begin{align*}
(b_0)_\text{cell} &= \frac{FkT_0}{P_0/m}
\end{align*}
at the output of each cell, and
\begin{align}
b_0 &= \frac{FkT_0}{P_0}
\label{eq:b0-PA}
\end{align}
at the output of the parallel amplifier, after adding $m$ signals of equal power, same statistical properties, and independent.  This also means that the noise figure of the parallel amplifier is equal to the noise figure $F$ of one cell.

Recalling that $S_{n''}$, and $S_{n'}$ as well, has $1/f$ spectrum,
the flicker noise can be derived from eq. (\ref{eq:phi-t}) applied
to one cell
\begin{align*}
(b_{-1})_\text{cell} & = 4\frac{a^2_2}{a^2_1}S_{n''}(\text{1\,Hz})~.
\end{align*}
Combining $m$ statistically-independent signals of equal power and same statistical properties gives
\begin{align*}
b_{-1} &= \frac{1}{m}\:4\frac{a^2_2}{a^2_1}S_{n''}(\text{1\,Hz})~,
\end{align*}
and therefore
\begin{equation}
b_{-1} = \frac{1}{m} (b_{-1})_\text{cell}
\label{eq:RatioPNoiseParallel}
\end{equation}

The important conclusion is that the parallel configuration features
a flicker-noise reduction of a factor $m$, or $3\log_2(m)$ dB,
assuming perfect symmetry and no dissipative losses in the
splitter/combiner networks.  This is shown in Fig.~\ref{fig:4schemes}\,D\@.
In practice, a noise reduction of 2.5 dB per factor-of-two is expected.  It is worth mentioning that
similar architectures have already been employed to reduce flicker
phase noise of photodiodes by connecting several units in parallel
\cite{SangSunLee2003jkps, Giboney1992ptl}.  Conversely, general theory
states that white noise cannot be improved in this way.  In
practice, the loss of the input power-splitter increases the noise figure, thus $b_0$.

\subsection{The regenerative amplifier}
\begin{figure}
\centering\includegraphics[scale=\SchemeScale]{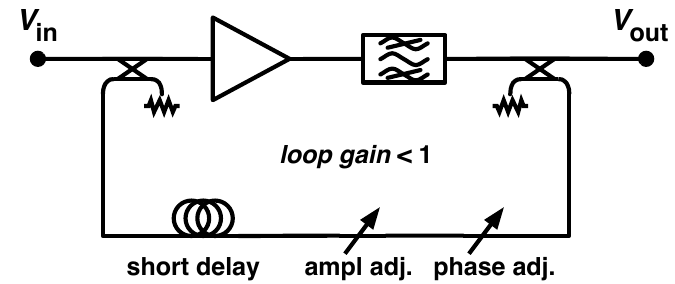}
\caption{Regenerative (positive-feedback) amplifier.}
\label{fig:RegenerativeAmp}
\end{figure}
The regenerative amplifier (RA) is an amplifier in which positive
feedback (regeneration) is used to increase the gain, as shown in
Fig.~\ref{fig:RegenerativeAmp}.  Interestingly, this technique is by
far better known in optics than in radio engineering.  A
sub-threshold laser is a common example of optical regenerative amplifier.

Denoting with $A_0$ the voltage gain of the simple amplifier and
with $\beta$ the gain of the feedback path, elementary feedback
theory suggests that the regenerative-amplifier gain is
\begin{align}
A &= \frac{A_0}{1-A_0\beta}~.
\label{eq:RegenGain}
\end{align}
As an analogy with $m$ cascaded amplifiers, we can also write
\begin{align}
A_0^m \qquad\text{with}\qquad
\beta=\frac{A_0^{m-1}-1}{A_0^m}~.
\label{eq:RegenGainRatio}
\end{align}
Of course, there is no reason to restrict this representation to integer $m$.

At a closer sight, one should introduce coupling coefficient
$\kappa_i$ and $\kappa_o$ of the input and output couplers, and also
the dissipative losses. Since the effect of the coefficient $\kappa$
is an intrinsic power loss $1-\kappa^2$ if the coupler, for the regenerative-amplifier
gain is reduced by a factor
$\smash{\sqrt{(1-\smash{\kappa^2_i})(1-\smash{\kappa^2_o})}}$.  The small effect of
the coupler losses will be neglected in the rest of this Section.

It is wise to adjust the phase for the roundtrip gain $A_0\beta$ to
be real, hence $G$ is real.  It is to be made sure that
$0<A_0\beta<1$.  The condition $A_0\beta>0$ means that the feedback
is positive, while $A_0\beta<1$ is necessary to keep the loop gain below the oscillation threshold.

The equivalent noise temperature is the noise
temperature of the internal amplifier referred to the RA input.  This is the temperature of
the internal amplifier increased by the loss of the input coupler.  The detailed analytical proof given in \cite{Brida-Q-multiplier} for the $Q$-multiplier, which is an application of the regenerative amplifier where a resonator is inserted in the feedback, holds for the regenerative amplifier in the general case.
The consequence is that the regenerative-amplifier white noise is
\begin{align}
b_{0} &= \frac{FkT_0}{P_0} + \text{losses}~.
\end{align}

The flicker noise is best understood by replacing the gain $A_0$ with
$A_0e^{j\psi}$, where $\psi(t)$ is the instantaneous value of the
internal-amplifier noise.  In practical design the flicker of phase
shows up at low frequencies, at least a factor of $10^2$ lower than
the inverse of the roundtrip time.  In these conditions the signal
circulating in the loop sees a quasi-static phase $\psi$, hence the
gain can be written as
\begin{align}
A &= \frac{A_0\,e^{j\psi}}{1-A_0\beta\,e^{j\psi}}
\end{align}
and expanded using $e^x=1+x$ for low noise
\begin{align}
A &= \frac{A_0}{1-A_0\beta} \left[
1+j\frac{1}{1-A_0\beta}\,\psi\right]~.
\end{align}
Accordingly the regenerative-amplifier (RA) phase noise is
\begin{align}
\varphi(t) &= \frac{1}{1-A_0\beta}\:\psi(t)
\end{align}
\begin{align}
(b_{-1})_\text{RA} &= \left[\frac{1}{1-A_0\beta}\right]^2
(b_{-1})_\text{ampli} \label{eq:b-1RA}~, \intertext{which after eq.
(\ref{eq:RegenGainRatio}) is equivalent to} (b_{-1})_\text{RA} &=
m^2(b_{-1})_\text{ampli}~.
\label{eq:b-1RAm}
\end{align}

It is instructive to compare the $1/f$ of a cascade of $m$
amplifiers to that of a regenerative amplifier.  The comparison makes sense only if the
two configurations use the same type of amplifier and have the same
gain.  The latter condition sets the value of $\beta$. It follows
from (\ref{eq:b-1ChainCasc}) that the flicker of the cascade is
\begin{align}
(b_{-1})_\text{chain} &= m(b_{-1})_\text{ampli}~,
\end{align}
thus
\begin{align}
(b_{-1})_\text{RA} &= m(b_{-1})_\text{chain}~.
\label{eq:regen-vs-cascade}
\end{align}
However counterintuitive, this conclusion not a surprise to us 
because in the chain the carrier is phase-shifted by $m$ independent
random processes, while in the regenerative-amplifier it is shifted $m$ times by the same slow process.

\subsection{The virtues of the error amplifier}
\begin{figure}
\centering\includegraphics[scale=\SchemeScale]{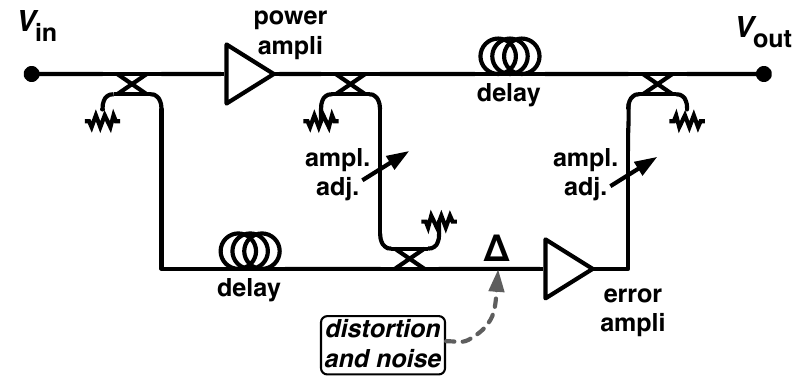}
\caption{Feedforward amplifier.}
\label{fig:FeedforwardAmp}
\end{figure}
A side effect of Eq.~(\ref{eq:b-1}) is that the amplifier noise sidebands are proportional to the carrier. Since the amplifier $1/f$ noise sidebands are proportional to the carrier, an error amplifier
that receives the null signal of a bridge is virtually free from close-in flicker.

The feedforward amplifier (Fig.~\ref{fig:FeedforwardAmp}, and Ref.~\cite{pothecary:feedforward}) is based on the idea that a low distortion is achieved by introducing an error amplifier that processes only the error of the power amplifier, which is a small signal.  For the same reasons, the feedforward amplifier also exhibits low flicker.   A review oriented to low phase noise applications is given in \cite{McNeilage1998fcs}.

\begin{figure}
\centering\includegraphics[scale=\SchemeScale]{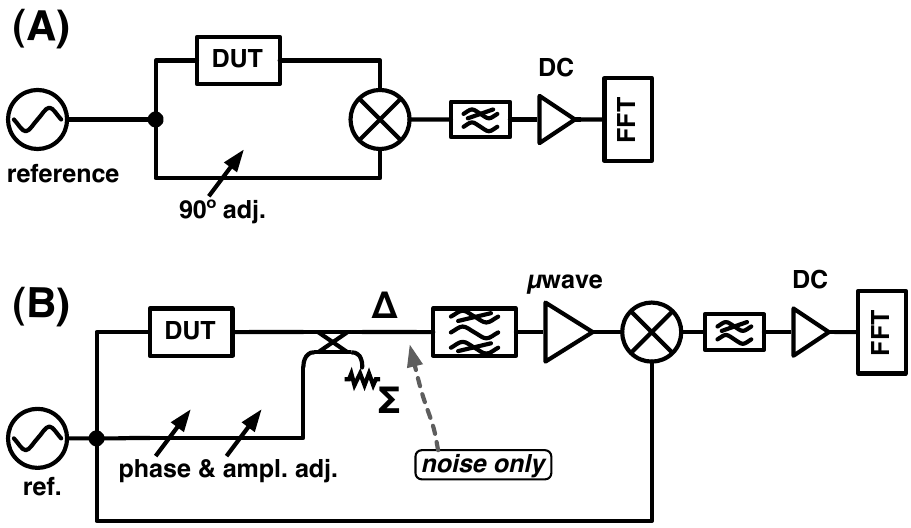}
\caption{Phase noise measurement methods. A: saturated mixer.  B:
low-flicker carrier-suppression scheme.} \label{fig:NMS}
\end{figure}

Our noise-measurement system of Fig.~\ref{fig:NMS}(B) exploits the fact that the amplifier cannot up-convert the near-dc $1/f$ noise if the  carrier is suppressed at its input.
More precisely, the contribution of the error amplifier to the background $b_{-1}$ is divided by the carrier rejection ratio, that is, approximately the DUT power divided by the residual carrier at the input
of the error amplifier.  This ratio can be of 60--100 dB\@.

If the device under test (DUT) of Fig.~\ref{fig:NMS}(B) is an amplifier shared by the
noise-measurement system and by an external circuit, we can use the noise-measurement system to null the amplifier noise in closed loop.  In practice, the $1/f$ noise is
limited by the background of the noise-measurement system\@.  This is used for the
reduction of the oscillator $1/f$ frequency noise
\cite{Dick1991patent,Ivanov2006mtt}, with detectors conceptually
equivalent to that of Fig.~\ref{fig:NMS}(B).

\subsection{The effect of physical size}
Physical insight suggests that the flicker coefficient $b_{-1}$ is
proportional to the inverse of the volume of the amplifier active
region.  This can be seen through a gedankenexperiment in which we
set up a $m$-cell parallel amplifier, whose flicker is
$b_{-1}=\frac1m(b_{-1})_\text{cell}$
[Eq.~\ref{eq:RatioPNoiseParallel}].  Then we join the $m$ cells
forming a single large device trusting the fact that flicker is of
microscopic origin and that the elementary volumes are uncorrelated.
This assumption is supported by the observation that the variety of flicker
models for specific cases share the fact that flicker is of
microscopic origin.  Moreover, the sum of a large number of
independent processes by virtue of the central-limit theorem yields
a Gaussian distribution, which is generally observed.

Our inverse-volume law must be taken with prudence.  First, for a
given volume flicker depends on technology.  Second, the volume law
certainly breaks down at nanoscale, where the size is smaller than
the coherence length of the flicker phenomenon and the elementary
volumes are no longer independent; and likely also at large scale.
Nonetheless, the inverse-volume law is a useful design guideline.

\begin{table}
\begin{center}
\caption{RF and microwave amplifiers tested.}\label{tab:CaracAmplis} \vspace*{1ex}
\begin{tabular}{|c|c|c|c|c|c|c|}
\hline
Amplifier&Frequency& Gain& $P_\text{1\,dB}$& $F$ & DC & $b_{-1}$ (meas.)\\
              &(GHz) &(dB) &(dBm) & (dB) & bias &(dBrad$^2$/Hz)\\
\hline
AML812PNB1901 & 8 -- 12 & 22 & 17 & 7 & 15\,V, 425\,mA & $-122$\\
AML412L2001 & 4 -- 12 & 20 & 10 & 2.5 & 15\,V, 100\,mA & $-112.5$\\
AML612L2201 & 6 -- 12 & 22 & 10 & 2 & 15\,V, 100\,mA & $-115.5$\\
AML812PNB2401 &  8 -- 12 & 24 & 26 & 7 & 15\,V, 1.1A & $-119$\\
AFS6 & 8 -- 12 & 44 & 16 & 1.2 & 15\,V, 171\,mA & $-105$\\
JS2 & 8 -- 12 & 17.5 & 13.5 & 1.3 & 15\,V, 92\,mA & $-106$\\
SiGe LPNT32 & 3.5 & 13 & 11 & 1 & 2\,V, 10\,mA & $-130$\\
Avantek UC573 & 0.01 -- 0.5 & 14.5 & 13 & 3.5 & 15\,V, 100\,mA & $-141.5$\\
\hline
\end{tabular}
\end{center}
\end{table}

\begin{figure}
\centering\includegraphics[scale=\SpectraScale]{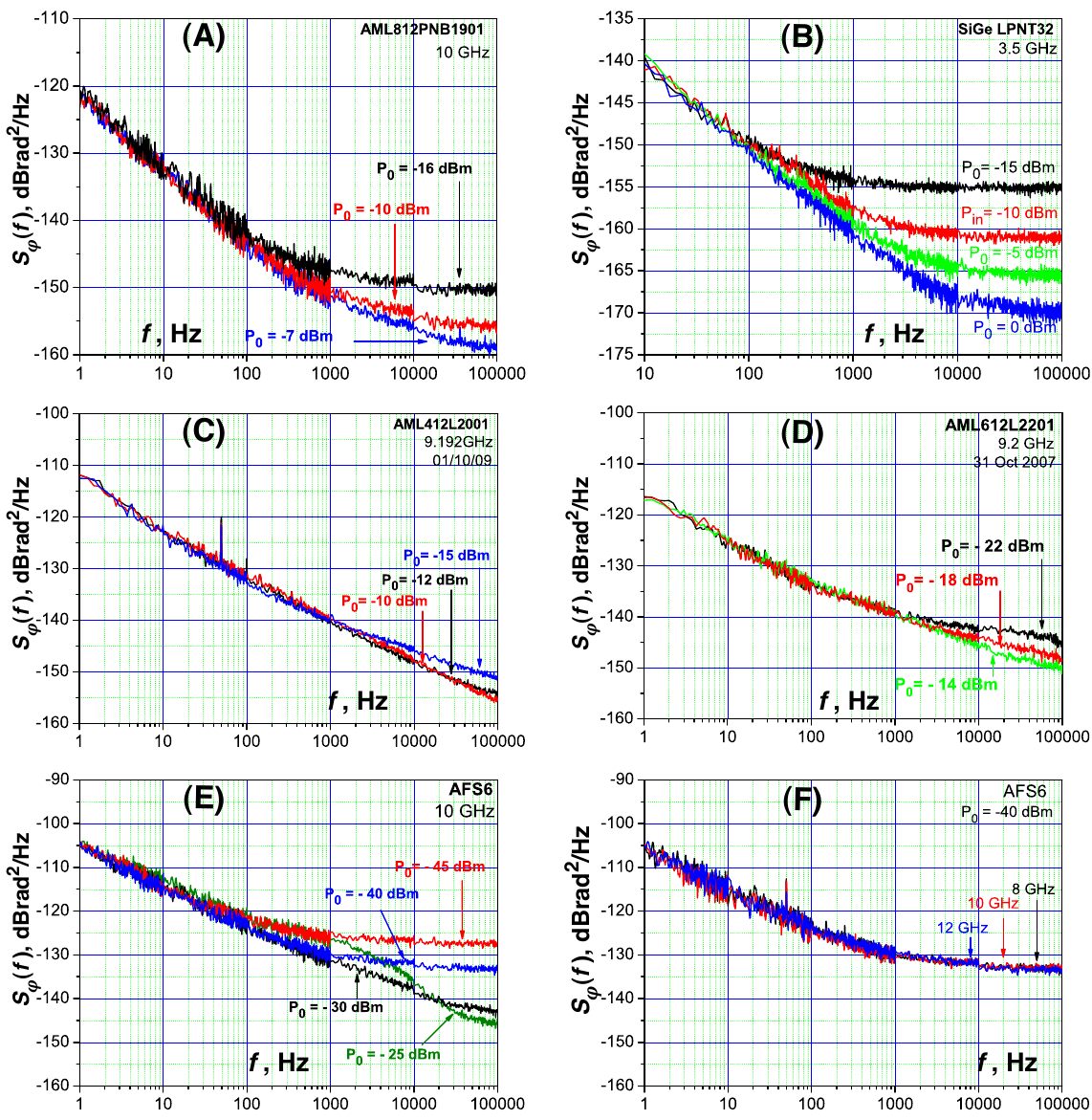}
\caption{Phase noise of some
amplifiers, measured at different input power and frequency.  The
plot (B) was measured at LAAS
(Toulouse, France) using the system described in
\cite{Cibiel2004mtt-Si-transistors}, and first made available in \cite{Boudot-2006-PhD} (Fig.~3.16).} 
\label{fig:NoiseVsPin}
\end{figure}

\section{Experimental Proof}\label{sec:ExpProof}
\subsection{Measurement method}\label{sec:MeasMethod}
Two different schemes, shown in Fig.~\ref{fig:NMS}, have been used
to measure the amplifier phase noise, depending on needs.  The
scheme A is that of commercial phase-noise measurement systems.  A
Schottky-diode double-balanced mixer saturated at  both inputs with
7--10 dBm driving power is used as the phase detector.  The two
inputs are to be in quadrature.  In this condition the mixer
converts the phase difference $\varphi$ into a voltage
$V=k_d\varphi$ with a typical conversion factor of 100--500 mV/rad.
The mixer output is low-pass filtered, amplified and sent to the fast Fourier transform (FFT)
analyzer. The background $1/f$ noise is chiefly due the mixer.
Typical values are of $-140$ dBrad$^{2}$/Hz for RF mixers and $-120$
dBrad$^{2}$/Hz for microwave mixers.  The white part of the
background noise is generally due to the dc-amplifier (1.5
nV/$\sqrt{\mathrm{Hz}}$) referred to the mixer input.  Values of $-155$ to
$-170$ dBrad$^{2}$/Hz are common in average or good experimental
conditions. The amplifier described in
\cite{Rubiola2004rsi-amplifier} is designed for this type of
applications, and optimized for the lowest background flicker when
connected to a 50 $\Omega$ source.

SiGe amplifiers may exhibit outstanding low $1/f$ noise, below the
instrument background.  When this happens, the scheme of
Fig.~\ref{fig:NMS}\,A is replaced with that of
Fig.~\ref{fig:NMS}\,B.  This detector, well known in the literature
\cite{Labaar-1982-Microwaves, Ivanov1998uffc, Rubiola1999rsi,
Rubiola2002rsi-matrix}, works as a Wheatstone bridge followed by a
microwave amplifier and a synchronous detector.  Since all the DUT
noise is contained in the sidebands, low $1/f$ background is
achieved by suppressing the carrier at the input of the microwave
amplifier labeled G\@.  The latter amplifies only the DUT noise
sidebands, which are low-power signal, so that virtually no flicker
up-conversion takes place.  Microwave amplification before detecting
has the additional advantage of low white background and reduction
of 50--60 Hz spurs.  This happens because the dc amplifier take in
low-frequency magnetic fields, while microwave amplifiers do not.
Neglecting dissipative losses, the white background is
\begin{equation}
(b_0)_\text{bg} = \frac{2FkT_0}{P_\text{hyb}}
\label{eq:sphi0_Bridge}
\end{equation}
where $F$ is the noise figure of the amplifier labeled G,
$P_\text{hyb}$ is the microwave power at the inputs of the hybrid
junction, and the factor 2 is the junction intrinsic loss. The value
of $-185$ dBrad$^{2}$/Hz is easily achieved at 15 dBm power level.
The $1/f$ background is not limited by necessary and known factors.
We obtained $(b_{-1})_\text{bg}=-150$ dBrad$^{2}$/Hz in the very
first experiments \cite{Rubiola1999rsi}, and
$(b_{-1})_\text{bg}=-180$ dBrad$^{2}$/Hz with a series of tricks
\cite{Rubiola2002rsi-matrix}.   The phase-to-voltage gain can be 40
dB higher than that of the saturated mixer. Interestingly, the
scheme of Fig.~\ref{fig:NMS}\,B can be built around a commercial
instrument (Fig.~\ref{fig:NMS}\,A), re-using mixer, dc amplifier,
FFT and data acquisition system. The only problem with
Fig.~\ref{fig:NMS}\,B is that the carrier suppression must be
adjusted manually, which may take patience, experimental skill, and
often replacing some parts when frequency is changed.

\subsection{Experimental results}
We measured the amplifiers listed in Table~\ref{tab:CaracAmplis}.
All are commercial products but the LPNT32, which was designed and
implemented at the Laboratoire d'Analyse et d'Architecture des Syst\`{e}mes (LAAS), Toulouse \cite{Cibiel-2004-UFFC--sapphire}. We
believe that the AML812PNB1901 and the AML812PNB2401, claimed to be
ultra-low noise units by AML, are actually parallel amplifiers.  The reason is that there is a series of five AML amplifiers with DC bias current in powers of two, from 0.1 to 1.6 A, and output power proportional to the DC bias.  Interestingly, $b _{-1}$ scales down by almost
3 dB per factor-of-two increase in the dc bias \cite[Chapter
2]{Rubiola-2008-Cambridge--Leeson-effect}. Our measurements aim at the
knowledge of the coefficient $b_{-1}$, and at the experimental
confirmation of the behavioral rules stated in
Section~\ref{sec:Rules}. The results are given as a series of
spectra discussed underneath.  Additionally, $b_{-1}$ is reported on
the right-hand column of Table~\ref{tab:CaracAmplis}.

White phase noise, though understood in the literature, is a
necessary complement to this work and a sanity check for the
results.

\subsubsection{Phase noise of a single amplifier}
The first experiment is the simple measurement of the phase noise of
several microwave amplifiers at different values of input power
(Figure \ref{fig:NoiseVsPin}).  It is clearly seen on all spectra
that $b_{-1}$ is independent of power.  The fact that $b_{-1}$ is constant vs power holds for different technologies, and in moderate compression regime.
This confirms the parametric nature of flicker and validates the main point of the behavioral model.

In Fig.~\ref{fig:NoiseVsPin}\,(A-B), 
the white noise $b_0$ follows exactly the $1/P_0$ law (\ref{eq:b0}).
The white phase noise cannot be observed in Figs.~\ref{fig:NoiseVsPin}\,(C-D)
because the frequency span of our FFT analyzer is insufficient.  
In Fig.~\ref{fig:NoiseVsPin}\,(E), 
the white noise $b_0$ follows exactly the $1/P_0$ law up to $-30$ dBm input power.  
At $-25$ dBm (dark green curve), we observe that between 100 Hz and 10 kHz the noise is higher than the flicker we expected from the general rules stated.  This is likely the consequence of saturation in an intermediate stage.

The  AML812PNB1901 and the LPNT32 [Fig.~\ref{fig:NoiseVsPin}\,(A-B)]
are intended for low phase noise applications and for high spectral purity oscillators \cite{Cibiel-2004-UFFC--sapphire, Boudot-2006-ELL--Low-noise-oscillator, Boudot-2006-PhD}.  These amplifiers exhibit $b_{-1}<120$ dBrad$^{2}$/Hz.  The white noise shown, though remarkably low, is the noise predicted by (\ref{eq:b0}).

It is worth mentioning that the power efficiency (output power divided by dc-bias power) is of 50\% for the LPNT32 (LAAS laboratory design \cite{Cibiel-2004-UFFC--sapphire}), and of 0.5\%--2.5\% for the
commercial amplifiers.  This indicates that low flicker design is not incompatible with efficiency.

Our experience indicates that the flicker of a given amplifier does not in the frequency range.
Since this fact is observed all the time, we did not repeat the test systematically end we show only one case in Fig.~\ref{fig:NoiseVsPin}\,(F).

\subsubsection{Cascaded amplifiers}
\begin{figure}[t]
\centering\includegraphics[scale=\SpectraScale]{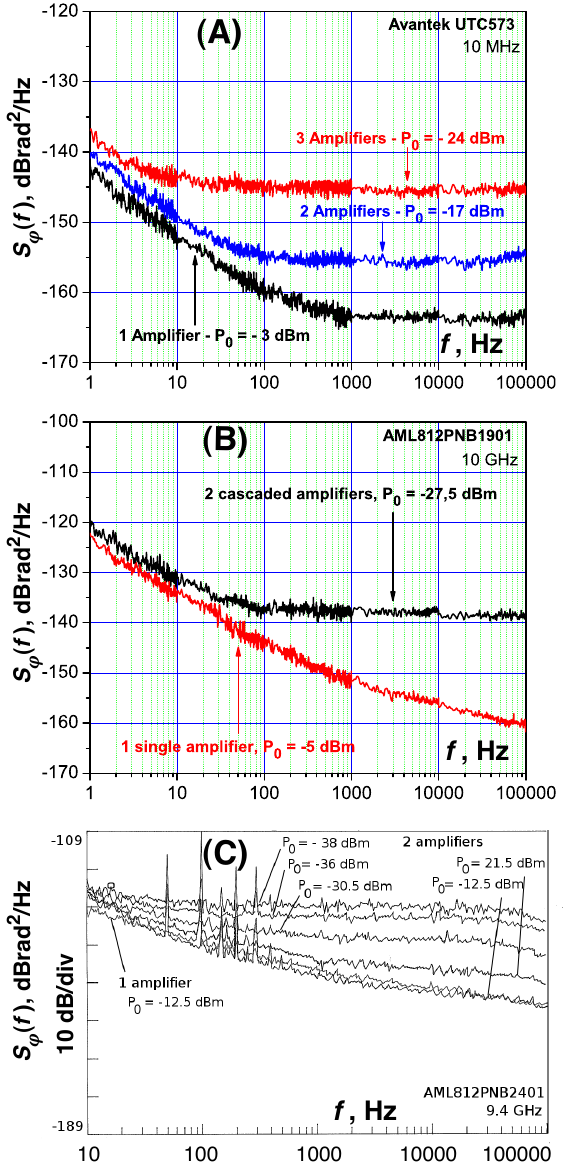}
\caption{Phase noise of
cascaded amplifiers, compared to the noise of a single amplifier.}
\label{fig:Cascade}
\end{figure}
In a second experiment we check on the rule of cascaded amplifiers versus Eq.~(\ref{eq:b-1ChainCasc}) by connecting 2--3 equal units.  We did not insert attenuators in the chain.  The consequence is that the input power must be scaled down proportionally to the total gain for the output to be kept in the linear or moderate-compression region. Yet, impedance matching is improved with microwave isolators.  The noise spectra are shown in Fig.~\ref{fig:Cascade}.

Figure~\ref{fig:Cascade}\,(A)
shows the phase noise of a chain consisting of 1--3 UTC573 operated at 10 MHz.  The flicker fits almost exactly the model, which predicts an increase of 3 dB for 2 cascaded units, and of 4.8 dB for 3 units.  The small discrepancy is ascribed to the difference between the amplifiers. The reference (one amplifier) is the noise of a single device instead of the average of the 2--3 amplifiers. For the single
amplifier measured at $-3$ dBm input power, the white noise hits the background of the instrument.  Otherwise it follows Eq.~(\ref{eq:b0chain}).
The same result is obtained with two AML812PNB1901 tested at 10 GHz, as seen on Fig.~\ref{fig:Cascade}\,(B).

Figure~\ref{fig:Cascade}\,(C)
shows the phase noise of two cascaded AML812PNB2401 at 10 GHz, measured at low input power and
compared to the single amplifier.  The flicker coefficient is $b_{-1}=-119$ dBrad$^{2}$/Hz for one amplifier, and $-116.5$ dBrad$^{2}$/Hz for the two amplifiers, independent of power. The reason for careful noise investigation in the microwatt range is that this amplifier is an important piece of the
frequency-synthesis chain used at SYRTE, Paris, for fundamental metrology \cite{Millo2009ol}. 
The amplifier receives a $-30$ dBm microwave signal from the output of a photodetector driven by a 250 MHz frequency comb.  At some point, the amplifier was suspected to flicker more than expected when used at low power.  

\subsubsection{Parallel amplifiers}
\begin{figure}
\centering\includegraphics[scale=\SpectraScale]{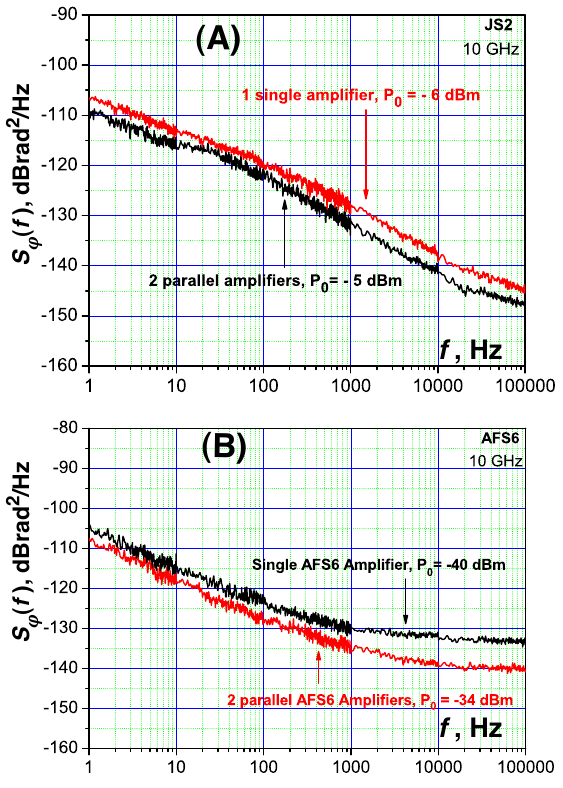}
\caption{Phase noise of parallel amplifiers.}
\label{fig:Parallel}
\end{figure}

In a third experiment we measured the phase noise of a pair of amplifiers (AFS6 or JS2) connected in parallel.  We used Wilkinson power splitters/combiners at the input and at the output instead of
$90^\circ$ couplers for the trivial reason that layout and trimming are simpler.  Anyway, the demonstration of our ideas is independent of the impedance-matching benefit of the $90^\circ$ couplers. The power $P_0$ refers to the main input, before splitting the signal. Measuring the AFS6, we had to adapt the power to experimental needs, while the JS2 could be measured at about the same level for the single amplifier and for the parallel configuration.
The spectra are shown in Fig.~\ref{fig:Parallel}. 

We observe that the flicker of the pair is 2.5 dB lower than the noise of the single amplifier, while the model predicts 3 dB\@.  This is ascribed to the gain asymmetry and to the asymmetry of the power splitter and combiner.

In Figure \ref{fig:Parallel}\,(A) 
we observe a significant discrepancy with respect to the power-law (\ref{eq:specAmpli}).  A slope of $-7$ dB/decade shows up in the left-hand side of the spectrum, up to 10--30 Hz, followed by a small bump.  Careful check indicates that there is no damage, and the result is reproducible.  
Having no explanation for this anomalous behavior, we report the spectrum as a counter example, yet the only one found.

\subsubsection{Regenerative amplifier}
\begin{figure}\centering
\includegraphics[scale=\OeoSpectrumScale]{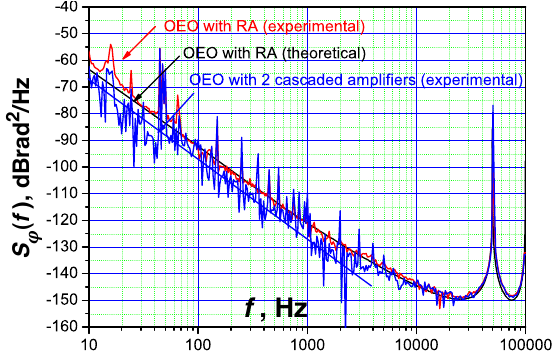}
\caption{Phase noise of an opto-electronic oscillator (OEO) set at 10 GHz carrier, from \cite{Volyanskiy2009phd}.}
\label{fig:OEO_RA_Kirill}
\end{figure}

The fourth experiment is the indirect measurement of the noise of a regenerative amplifier in the loop of an oscillator.  According to the \emph{Leeson effect}, the oscillator integrates the phase noise of the sustaining amplifier \cite[Chapter 4]{Rubiola-2008-Cambridge--Leeson-effect}.  Hence, the flicker noise $(b_{-1})_\text{ampli}/f$ of the amplifier is transformed into frequency flicker, which shows up as a term $(b_{-3})_\text{osc}/f^3$ in the oscillator spectrum.  The oscillator $(b_{-3})_\text{osc}$ is a constant that can be calculated from $(b_{-1})_\text{ampli}$ and the resonator relaxation time.  
In the case of the delay-line oscillator \cite{Volyanskiy-2008-JOSAB--Optical-delay-line}, \cite[Chapter 5]{Rubiola-2008-Cambridge--Leeson-effect}, the noise transformation takes the form  $(b_{-3})_\text{osc}=\smash{\frac{1}{4\pi^2\tau^2}}(b_{-1})_\text{ampli}$, where $\tau$ is the roundtrip delay (20 $\mu$s in our case) of the oscillator loop.

In quite a different research program,  a colleague was investigating on high-spectral-purity photonic oscillators in which the microwave frequency is set by the delay $\tau$ of an optical fiber in the oscillator loop \cite{Yao1996josab-oeo, Volyanskiy-2008-JOSAB--Optical-delay-line, Volyanskiy2009phd}.  At some point he used regeneration to ``double'' (in dB) the gain of an AML812PNB1901 as a temporary  replacement for two cascaded amplifiers, and eventually he exploited the frequency response that derives from regeneration as a bandpass filter.  Then he asked for help in the interpretation of the measured noise, unfortunately, without having recorded the noise of the regenerative amplifier alone.

Figure \ref{fig:OEO_RA_Kirill} shows two oscillator spectra, one with a regenerative amplifier used to obtain 44 dB gain from one 22 dB AML amplifier, and the other with two amplifiers of the same type, cascaded.  Knowing the $1/f$ noise of the AML812PNB1901, we calculate the oscillator $1/f^3$ noise for the two cases.  The results (Fig.~\ref{fig:OEO_RA_Kirill}) are in a close agreement with the theory.    In the $1/f^3$ region ($10^1$--$10^3$ Hz), the noise is 3 dB higher when the regenerative amplifier is used instead of the two cascaded amplifiers, as expected from Eq.~(\ref{eq:b-1RAm}) and (\ref{eq:regen-vs-cascade}).  This fact validates the model.

\section{Final remarks}\label{sec:After-fact}
\begin{figure}
\centering\includegraphics[scale=\GeneralizedModelScale]{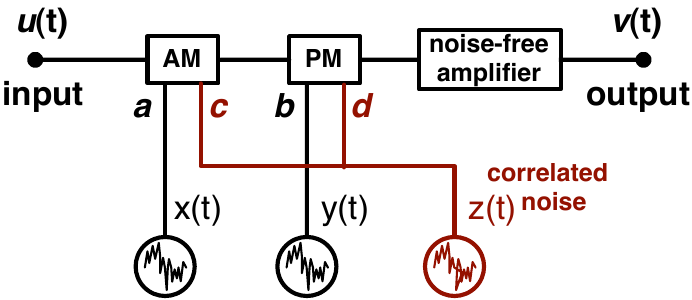}
\caption{Generalized model for the AM-PM noise in amplifiers.}
\label{fig:noise-correl-1}
\end{figure}

\begin{figure}
\centering\includegraphics[scale=\SimulationScale]{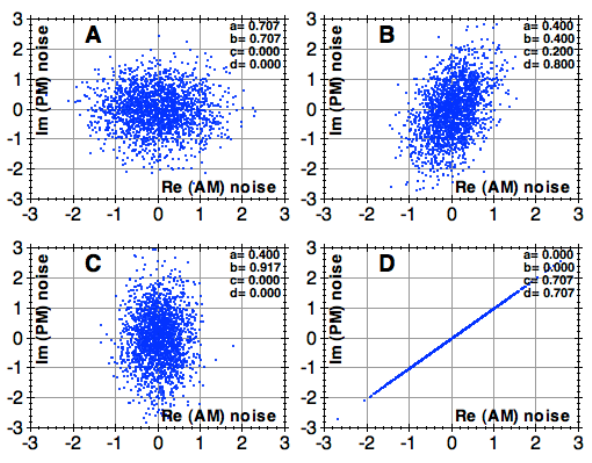}
\caption{Simulated parametric noise, real part (AM noise) and imaginary part (PM noise).  The
coefficient $a$, $b$, $c$, $d$ are defined in Fig.~\ref{fig:noise-correl-1}.}
\label{fig:noise-correl-2}
\end{figure}

This work  derives from a long-term research program on high-end oscillators and on frequency synthesis mainly for metrology and for military and space applications.  The measurements reported here were done in different contexts, over more than five years.  
In the domain of oscillators, people are interested only in PM noise, while AM noise is considered a scientific curiosity and mentioned only for completeness. Amplitude noise is sometimes measured carefully \cite{rubiola2005arxiv-am-noise}, yet for quite different purposes, or is investigated because of its detrimental effect on phase-noise measurements \cite{Rubiola-2007-UFFC--AM-2-PM-pollution}.

It was only at the time of writing that evidence popped into our mind, that parametric AM and PM noise processes are partially correlated, and therefore that the amplifier noise is best modeled as in Fig.~\ref{fig:noise-correl-1}. 
The necessity for this model is justified by the physics of the most popular amplifier devices.  
In a bipolar transistor, the fluctuation of the carriers in the base region acts on the base thickness, thus on the gain and on the capacitance of the reverse-biased base-collector junction.  Of course, a fluctuating capacitance impacts on phase noise. 
In a field-effect transistor, the fluctuation of the carriers in the channel acts on the drain-source current, thus on the gate-channel capacitance via the channel thickness. 
In a laser amplifier, the fluctuation of the pump power acts on the density of the excited atoms, and in turn on gain, maximum power, and refraction index.  
In all these examples AM and PM fluctuations are correlated because both originate from a single near-dc random process.

Since the noise measurements are now terminated or put on hold, and instruments and components scattered in the lab, we can only support the model with simulations .  In the simulations shown in Fig.~\ref{fig:noise-correl-2}, we normalize on the carrier power, we linearize for low noise, and we set $a^2+b^2+c^2+d^2=1$ so that the noise power is equal to one.  
The simulated noise is shown as it would be measured by the two-channel version of the noise-measurement system shown in Fig.~\ref{fig:NMS}(B), where we detect simultaneously the real and the imaginary part with a I-Q mixer \cite{Rubiola2002rsi-matrix}.

In simplest form, the noise is a Gaussian process of power equally split into the real and imaginary part.  This is the symmetric two-dimensional Gaussian distribution of Fig.~\ref{fig:noise-correl-2}(A).
If the noise is not equally split between AM and PM, as for example it happens when the amplifier is the power compression region, there results an asymmetric Gaussian distribution
(Fig.~\ref{fig:noise-correl-2}(C)).
The perfectly saturated  amplifier has no AM noise, so it would be represented as a vertical line in a scatter plot.

Fig.~\ref{fig:noise-correl-2}(B) shows the case of flicker noise of
an amplifier operated in the compression region. The amount of AM
and PM is not the same, but there is some correlation between AM and
PM noise.  For comparison, the plot of Fig.~\ref{fig:noise-correl-2}(D) represents a (unrealistic)
amplifier in which AM and PM noise originates from a single random
process with the same modulation efficiency.

\section*{Acknowlegments}\label{sec:Acknow}
We thank Yannick Gruson for help with phase noise
measurements, and Vincent Giordano for support
and discussions.

\def\bibfile#1{/Users/rubiola/Documents/Articles/Bibliography/#1}
\addcontentsline{toc}{section}{References}
\bibliographystyle{ieeetr}
\bibliography{\bibfile{ref-short},\bibfile{references},\bibfile{rubiola}}

\end{document}